\documentclass[aps,twocolumn,prd,epsf]{revtex4}
\usepackage{amsmath,amssymb}
\usepackage{graphicx}

\def\beq{\begin{equation}}
\def\eeq{\end{equation}}
\def\ba{\begin{eqnarray} }
\def\ea{\end{eqnarray}}

\begin{document}

\title{Non-commutative inflation and the CMB}

\author{Shinji Tsujikawa$^1$,
Roy Maartens$^1$, and Robert Brandenberger$^2$ }

\address{~}
\address{$^1$Institute of Cosmology and Gravitation,
University of Portsmouth, Portsmouth PO1~2EG, UK}

\address{$^2$Physics Department, Brown University,
Providence RI 02912, USA}

\date{\today}

\begin{abstract}

Non-commutative inflation is a modification of standard general
relativity inflation which takes into account some effects of the
space-time uncertainty principle motivated by ideas from string
theory. The corrections to the primordial power spectrum which
arise in a model of power-law inflation lead to a suppression of
power on large scales, and produce a spectral index that is blue
on large scales and red on small scales. This suppression and
running of the spectral index are not imposed ad hoc, but arise
from an early-Universe stringy phenomenology. We show that it can
account for some loss of power on the largest scales that may be
indicated by recent WMAP data. Cosmic microwave background
anisotropies carry a signature of these very early Universe
corrections, and can be used to place constraints on the
parameters appearing in the non-commutative model. Applying a
likelihood analysis to the WMAP data, we find the best-fit value
for the critical wavenumber $k_*$ (which involves the string
scale) and for the exponent $p$ (which determines the power-law
inflationary expansion). The best-fit value corresponds to a
string length of $L_s \sim 10^{-28}\,{\rm cm}$.

\end{abstract}


\maketitle

\section{Introduction}

General relativity will break down at very high energies in the
early Universe when quantum effects are expected to become
important. If the very early Universe is described by a period of
inflation~\cite{Guth}, as in the current paradigm of cosmology,
and if the period of inflation lasts sufficiently long (as it does
in most current scalar field-driven inflationary models), then,
since the wavelength of perturbations which are observed today
emerged from the Planck region in the early stages of inflation,
in principle the quantum theory of gravity should leave an imprint
on the primordial spectrum of perturbations (see~\cite{initial}
for the first discussion of this effect; see~\cite{review} for a
recent review containing a comprehensive list of references,
and~\cite{recent} for some of the latest papers on this issue).
The nature of these imprints will remain an open question as long
as we lack a complete theory of quantum gravity. If we assume that
string theory is a promising framework for quantum gravity, it is
of interest to explore specific stringy corrections to the
spectrum of fluctuations.

The standard concordance $\Lambda$CDM model, which can arise from
an inflationary background cosmology in which the
quasi-exponential expansion of space is driven by a scalar field,
provides a good fit to the recent
WMAP~\cite{Spergel:2003cb,Kogut:2003et,Verde:2003ey,Peiris:2003ff}
and earlier observations. This implies, in particular, that any
stringy corrections to general relativity will be constrained by
the properties of the observed cosmic microwave background (CMB)
anisotropies. Although there is no signature in CMB data of
statistically significant deviations from the predictions of the
standard paradigm, the unexpectedly low quadrupole and
octopole~\cite{Spergel:2003cb} are intriguing, in particular since
a similar deficit of power on these large angular scales was also
seen in the earlier COBE maps. Thus, although the lack of power on
the largest scales may simply be a statistical effect (and
different approaches to statistical analysis yield differing
results concerning the statistical significance of the lack of
power~\cite{ge}), early Universe explanations of this lack of
power are not ruled out and are worth exploring, provided that the
explanations are not ad hoc.

Recent papers~\cite{low1,low2,Contaldi,Cline,Feng} have attempted
to explain the low quadrupole and octopole, typically via a
finely-tuned inflationary potential or an ad hoc cut-off in the
spectrum (see Ref.~\cite{Yokoyama} for the earlier work
about this suppression).
In Ref.~\cite{KYY}, supergravity inflation is shown to
lead to a possible running of the power spectrum which tends to be
blue on large scales.  In Ref.~\cite{bfm}, some corrections to
general relativity inflation motivated by brane-world scenarios
(correction terms in the Friedmann equations and
velocity-dependent potentials) were studied, and it was shown that
they may account for the suppression of power. Here, we
investigate consequences of a more basic stringy effect, namely
the space-time uncertainty relation~\cite{uncpr}
\beq
\label{uncrel}
\Delta t \Delta x_{\rm phys} \, \geq \, L_s^2 \,,
\eeq
where $t,x_{\rm phys}$ are the physical space-time coordinates and
$L_s$ is the string scale. In~\cite{Alexander:2001dr}, it was
shown that this principle may yield inflation from pure radiation.
A more modest approach was pioneered
in~\cite{Brandenberger:2002nq}, where the consequences were
studied of imposing Eq.~(\ref{uncrel}) on the action for
cosmological perturbations on an inflationary background
(generated in the usual way).

Space-time non-commutativity~\cite{excep} at high energies in the
early Universe leads to a coupling between the fluctuations
generated in inflation and the background Friedmann model that is
nonlocal in time. In addition, the uncertainty relation is
saturated for a particular comoving wavelength when the
corresponding physical wavelength is equal to the string length.
Thus, fluctuation modes must be considered to emerge at this time
(a time denoted by $\tau_k$ later in the text),
and the most conservative assumption is to start the modes in the
state which minimizes the canonical Hamiltonian at that time. In a
background space-time with power-law inflation, these
modifications lead to a suppression of power for large-wavelength
modes (those created when the Hubble constant is largest),
compared to the predictions of standard general relativity
inflation~\cite{com1}.
It may appear counter-intuitive
that high-energy stringy effects modify
the large-scale perturbations rather than those on small scales.
But the point is that large-scale modes, which correspond
to higher energies earlier in inflation, are created outside the
Hubble radius due to stringy effects,
and thus experience less growth than the small-scale modes, which are
created inside the Hubble radius at lower energies, and evolve
as in the standard case.
There is a critical wavenumber $k_{{\rm crit}}$ such that for $k <
k_{\rm crit}$ the mode is created on super-Hubble scales, and thus
undergoes less squeezing during the subsequent evolution than it
does for $L_s = 0$. This critical wavenumber depends on the string
scale and on the exponent $p$ appearing in the formula for the
time-dependence of the scale factor [see Eq.~(\ref{sfactor})]. The
spectrum is blue-tilted for $k \ll k_{\rm crit}$
rather than red-tilted as it is in the power-law inflation
scenario with $L_s = 0$.

Here we calculate the spectrum of CMB anisotropies predicted by
the model of~\cite{Brandenberger:2002nq}, and thus quantify the
prediction of loss of power for infrared modes. In addition, we
perform a likelihood analysis to find the best-fit values to the
WMAP data of the cosmological parameters, including the power-law
exponent $p$ which gives the time dependence of the scale factor,
and the critical wavenumber $k_*$
(to a first approximation the same as $k_{\rm crit}$)
when stringy effects become
important. Thus we are able to constrain the non-commutative model
and place limits on the string scale. These results expand on the
previous work of~\cite{Huang:2003zp}, where the parameters of the
model of~\cite{Brandenberger:2002nq} were fitted to the WMAP data
at two specific angular scales.

\section{Non-commutative modifications
to the primordial power spectrum}

The stringy space-time uncertainty relation is compatible with an
unchanged homogeneous background, but it leads to changes in the
action for the metric fluctuations. The action for scalar metric
fluctuations can be reduced to the action of a real scalar field
$\phi$ with a specific time-dependent mass which depends on the
background cosmology (see e.g.~\cite{MFB} for a comprehensive
review). For simplicity, we will assume that matter is described
by a single real scalar field $\varphi$. In this case, the stringy
space-time uncertainty relation leads to the following modified
action for the field $\phi$~\cite{Brandenberger:2002nq}
\beq S \, = \, V_T \int
d\eta\,d^3k\,z_k^2(\eta)\left[\phi^\prime_{-k} \phi_k^\prime
-k^2\phi_{-k} \phi_k\right]\,, \label{action}
\eeq
where $V_T$ is the total spatial coordinate volume, a prime
denotes the derivative with respect to conformal time $\eta$, $k$
is the comoving wave number, and
\ba \label{zk2} && z_k^2= z^2 \left(\beta_k^{+}
\beta_k^{-}\right)^{1/2}\,,~ z={a\dot{\varphi} \over
H}\,,\label{s1}\\&& \beta_k^\pm ={1 \over 2} \left[a^{\pm
2}(\tau+kL_s^2)+a^{\pm 2}(\tau-kL_s^2) \right]\,, \label{zk}
\ea
where $a$ and $H=\dot{a}/a$ are the scale factor and the Hubble
rate, respectively, and $\tau$ denotes a new time variable
(related to the conformal time $\eta$ via $d\tau=a^2 d\eta$) in
terms of which the stringy uncertainty principle takes the simple
form $\Delta \tau \Delta x \geq L_s^2$, using comoving coordinates
$x$. The case of general relativity corresponds to
$L_s=0$. The nonlocal coupling in time between the
background and the fluctuations is manifest in Eq.~(\ref{zk}).

This stringy modification is complicated, and computing the power
spectrum will in general require numerical evaluation. However,
the spectrum can be evaluated analytically in power-law inflation,
and thus we assume for simplicity that
\beq \label{sfactor}
a(t) \, = \, a_0t^p \, ,
\eeq
with suitable
exponent $p>1$, is a reasonable approximation to the background
dynamics around the time when large-scale fluctuations are
generated.  Such a background can be obtained if inflation is
driven by a single scalar field with an exponential potential,
$V=V_0 \exp(-\sqrt{2/p}\,\varphi/M_{pl})$. In this case,
$\dot{\varphi}= {\sqrt{2p}M_{pl}/t}$, $H=p/t$ and
$z=a\sqrt{2/p}M_{pl}$ (here $M_{pl}$ is the reduced Planck mass,
$M_{pl}=2.4 \times 10^{18}\,{\rm GeV}$).

The power spectrum of the curvature perturbation, ${\cal R} \equiv
\Phi+H\delta\varphi/\dot{\varphi}$, where $\Phi$ is the scalar
metric perturbation in longitudinal gauge,
is~\cite{Brandenberger:2002nq}
\beq
\label{powerspe}
{\cal P}_{\cal R}={k^2\over
4\pi^2z_k^2(\tau_k)}\,,
\eeq
where $\tau_k$ is the time where the fluctuations are generated.

For $L_s \neq 0$, the power spectrum will have a different slope
for small values of $k$ than for large values. For large values,
one obtains the usual power spectrum with index $-2/(p -1)$,
whereas for very
small values of $k$, the spectrum is blue.
We have calculated approximations
to the power spectrum which become exact
either for very large or very small scales.

In~\cite{Huang:2003zp}, the power spectrum was calculated (up to
the normalization coefficient) by evaluating the time when the
fluctuations are generated. We have repeated the analysis in order
to compute the coefficient of the resulting spectrum (which yields
important information for placing limits on the string scale
$L_s$) in terms of $p$ and $k_0 \equiv a_0^{1/p}$. (This was not
done in~\cite{Huang:2003zp}.) We obtain
\beq {\cal P}_{\cal R} \simeq
A_1\left(\frac{k}{k_0}\right)^{-2/(p-1)}
\left[1-\left(\frac{k_{s1}}{k}\right)^{4/(p-1)}\right]\,,
\label{power2}
\eeq
where
\ba
 \label{amp}
 A_1 &=& \frac{p(2p^2-p)^{p/(p-1)}}{8\pi^2}\left(\frac{k_0}
 {M_{pl}}\right)^2\,,\\
k_{s1}& =& p^{{(3p-1)}/{4}}\sqrt{2p-1} \nonumber  \\
&&~{}\times [4(p-2)(2p+1)]^{{(p-1)}/{4}}  (k_0L_s)^p L_s^{-1}.
\label{cut} \ea

In deriving the relation (\ref{power2}), we made the assumption
that $k_{s1}\ll k$, and expanded everything in terms of
$k_{s1}/k$. Thus, the above power spectrum ceases to be valid as
$k$ approaches $k_{s1}$. For modes satisfying $k<k_{s1}$, we can
obtain the power spectrum by considering the fluctuations outside
the Hubble radius starting in a
vacuum~\cite{Brandenberger:2002nq}. According to the stringy
space-time uncertainty principle, we have an upper bound for the
comoving momentum,
\beq k_{\rm max}(\tau) \, = \, \left({\beta_k^+ \over
\beta_k^-}\right)^{1/4}L_s^{-1} \, . \eeq
By solving this equation one can find the initial time $\tau_k$ at
which the perturbation with comoving wavenumber $k$ is generated,
which yields \cite{Brandenberger:2002nq}
\beq
\tau_k=\left[k^2L_s^4+{(kL_s)^{2(p+1)/p}\over k_0^2(p+1)^2}
\right]^{1/2}.
\label{tauexact}
\eeq

The power spectrum of the curvature perturbation is derived by
inserting the time (\ref{tauexact}) into Eq.~(\ref{powerspe}),
but the general form of ${\cal P}_{\cal R}$ is very complicated.
When $\tau_k \gg kL_s^2$, which is the same approximation
used to derive Eq.~(\ref{power2}),
the power spectrum is given by
\beq
\label{powerspei}
{\cal P}_{\cal R} \simeq A_2
\left[1-\left(\frac{k_{s2}}{k}\right)^{2/p} \right]\,, \eeq
where
\ba
 \label{ampi}
A_2 &=& \frac{p}{8\pi^2}\left(\frac{L_{pl}}{L_s}\right)^2\,, \\
k_{s2} &=& [p(3p+1)]^{p/2} (k_0L_s)^p L_s^{-1}\,,
\label{cuti}
\ea
with $L_{pl} \equiv M_{pl}^{-1}$.
Note that the spectrum (\ref{powerspei})
is scale-invariant for $k \gg k_{s2}$.

Using Eqs.~(\ref{cut}) and (\ref{cuti}), one can easily show that
$k_{s2}$ is much smaller than $k_{s1}$ as long as the power-law
exponent $p$ satisfies $p \gg 1$. The two spectra,
Eqs.~(\ref{power2}) and (\ref{powerspei}), are joined at a value
of $k_*$ satisfying $k_* \gg k_{s1}$. Since $k_{s2} \ll k_{s1}$,
the second term on the right hand side of Eq.~(\ref{powerspei}) is
practically negligible on scales $k \gtrsim k_{s1}$.

The two cut-off scales, where the power spectra~(\ref{power2}) and
(\ref{powerspei}) formally vanish, satisfy $k_{s1} \gg k_{s2}$
for $p \gg 1$.  For example, when $p=20$ we have $k_{s1} \sim
10^6k_{s2}$.
We are interested in cosmologically relevant scales, which lie in the
range $10^{-4}\,{\rm Mpc}^{-1}<k<10^{-1}\,{\rm Mpc}^{-1}$.
When the power spectrum is mainly characterized by
Eq.~(\ref{power2}) on these scales, the spectral
index, $n=1+ {\rm d} \ln {\cal P}_{\cal R}/ {\rm d}\ln k$,
is given by
\beq n \, = \, 1-\frac{2}{p-1}\left[1-2 \left(\frac{k_{s1}}{k}
\right)^{4/(p-1)}\right]\,. \eeq
Then, the critical scale at which the spectrum becomes
scale-invariant is
\beq k_{n=1}=2^{(p-1)/4}\, k_{s1}\,. \eeq
The second term in the square bracket of Eq.~(\ref{power2}) is
$-1/2$ for $k=k_{n=1}$.  Therefore, the spectrum (\ref{power2}) is
not reliable for $k<k_{n=1}$ (corresponding to $n>1$), due to the
breakdown of the approximation, $(k_{s1}/k)^{4/(p-1)} \ll 1$.
This means that $k_{n=1}$ is not a suitable scale for joining the
spectral formulas~(\ref{power2}) and (\ref{powerspei}). Instead,
we need to join them at a scale $k_* > k_{n=1}$.

In~\cite{Huang:2003zp}, the likelihood values of $k_{s1}$ and $p$
are derived by using recent WMAP data on only two scales, namely
$n=0.93^{+0.03}_{-0.03}$, ${\rm d}n/{\rm d}{\rm ln}\,k=
-0.031^{+0.016}_{-0.017}$ at $k=0.05\,{\rm Mpc}^{-1}$ and
$n=1.10^{+0.07}_{-0.06}$, ${\rm d}n/{\rm d}{\rm
ln}\,k=-0.042^{+0.021}_{-0.020}$ at $k=0.002\,{\rm Mpc}^{-1}$.
However, we cannot trust the spectrum~(\ref{power2}) at the scale
$k=0.002\,{\rm Mpc^{-1}}$, since it is in the range where the
spectrum has a blue tilt ($n>1$).  In addition, the analysis
in~\cite{Huang:2003zp} makes use of the information only at the
above two scales, but it is not clear whether the fit will be good
on all scales.

\section{Non-commutative effects on the CMB, and likelihood analysis}

In order to compare the theoretical predictions of our class of
models with the recent WMAP results, we ran the CosmoMc
(Cosmological Monte Carlo) code (which makes use of the CAMB
program~\cite{CAMB}), developed in~\cite{antony}. This program
uses a Markov-chain Monte Carlo method to derive the likelihood
values of model parameters.  The method produces a large set of
sample spectra associated with given values of the model
parameters and of the usual cosmological parameters, and compares
them with the recent WMAP~\cite{Kogut:2003et,Spergel:2003cb} data
files from~\cite{WMAP} of temperature (TT) and
temperature-polarization cross-correlation (TE) anisotropy
spectra, by evaluating the $\chi^2$-distribution. We also include
the band-powers on smaller scales corresponding to $800 < l <
2000$, from the CBI~\cite{CBI}, VSA~\cite{VSA} and
ACBAR~\cite{ACBAR} experiments.  The contribution of gravitational
waves is also taken into account, since it
can be important for low multipoles.
Tensor perturbations can be obtained by replacing $z$ by $a$ in
the first of equations~(\ref{zk2}).  Taking into account the two
polarization states, we obtain the spectrum of
gravitational waves as
\beq {\cal P}_{T} = \frac{4}{p}\,{\cal P}_{\cal R} \,.
\label{powerT}
\eeq

\begin{figure}
\begin{center}
\includegraphics[width=8cm,height=7cm]{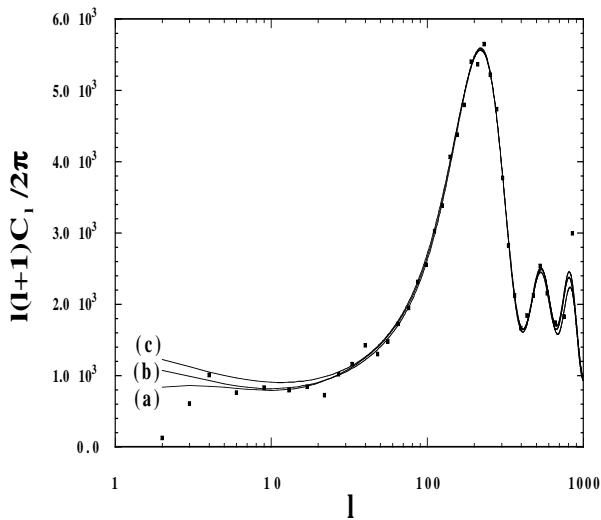}
\end{center}
\caption{ The CMB angular power spectrum showing the effects of
non-commutativity in the primordial spectrum
for power-law inflation. Curve~(a) corresponds to the best-fit
case where only the spectrum (\ref{power2}) is used for the
likelihood analysis. Curve~(b) shows the best-fit case in which
two power spectra (\ref{power2}) and (\ref{powerspei}) are
connected at $k_* \sim 10^{-2}\,{\rm Mpc}^{-1}$. Curve~(c) is the
standard $\Lambda$CDM model without non-commutative effects
($L_s=0$),
and for power-law inflation (i.e., without running of
the spectral index).} \label{TT}
\end{figure}

\begin{figure}
\begin{center}
\includegraphics[width=8cm,height=8cm]{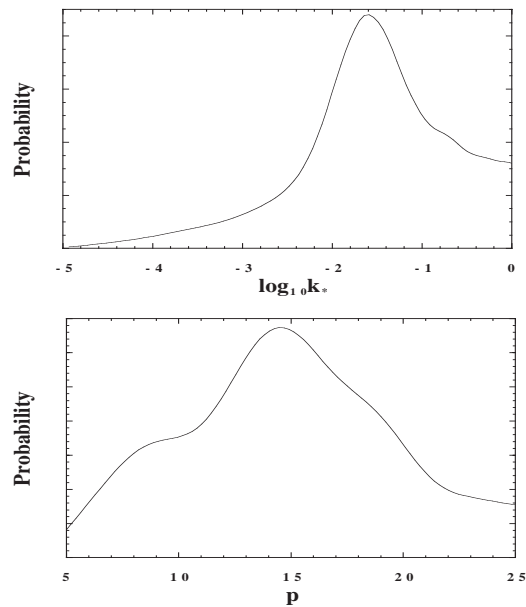}
\end{center}
\caption{Probability distributions of the mean likelihood analysis
for the parameters ${\rm log}_{10}\,k_*$ and $p$, when the two
power spectra (\ref{power2}) and (\ref{powerspei}) are joined at
$k=k_*$ (the units of $k_*$ on the horizontal axis are ${\rm
Mpc}^{-1}$, and the vertical axis is linear and ranges from 0 to
1). The probability distributions of other cosmological parameters
are not shown, but they basically agree with the ones in the
standard $\Lambda$CDM model.  } \label{noncomtens}
\end{figure}

Our results are summarized in Figs.~\ref{TT} and \ref{noncomtens}.
When only the spectrum (\ref{power2}) is used, the value of the
cut-off wavenumber with the highest likelihood is found to be
$k_{s1} \sim 10^{-4}\,{\rm Mpc}^{-1}$ .  In Fig.~\ref{TT} we show
the best-fit plot of the CMB temperature angular power spectrum
[see the case (a)].  In the absence of non-commutativity, i.e. for
$k_{s1}=0$, the best-fit value of $p$ is found to be $p \simeq
35$, thereby yielding a constant, red-tilted spectral index, $n
\simeq 0.94$.  In this context, it is difficult to explain the
suppressed power on large scales. On the other hand, space-time
non-commutativity can naturally lead to an effective running of
the spectral index.  In fact, as seen in Fig.~\ref{TT}, a better
fit for the quadrupole and octopole moments
can be obtained by taking into account the non-commutative effect.

However, the spectrum (\ref{power2}) is reliable only for scales
satisfying $k \gg k_{s1}$, implying that the prediction of the
curve (a) in Fig.~\ref{TT} will have some theoretical error for
values of $l$ in the range $2 \le l \lesssim 10$. In order for the
spectrum (\ref{power2}) to be valid for the cosmologically
relevant scales, $10^{-4}\,{\rm Mpc}^{-1}<k<10^{-1}\,{\rm
Mpc}^{-1}$, one needs to choose a smaller cut-off scale, $k_{s1}
\ll 10^{-4}\,{\rm Mpc}^{-1}$.  In this case, the deviation of the
CMB angular spectrum from what is obtained without a cut-off is
not significant ($k_{s1}=0$, results shown in (c) of
Fig.~\ref{TT}). Still, an analysis taking into account a spectrum
which on large scales is modified as compared to (\ref{power2}),
is required in order to avoid the problem of negative values of
${\cal P}_{\cal R}$ for $k<k_{s1}$, which occurs if we use the
spectrum (\ref{power2}) only.

To obtain a more complete analysis, it is necessary to match the
two spectra (\ref{power2}) and (\ref{powerspei}). Since $n=1$ at
$k=k_{n=1}$, it seems natural to consider that the spectrum
(\ref{power2}) is connected to (\ref{powerspei}) at $k_*=k_{n=1}$.
However, the approximation $(k_{s1}/k)^{4/(p-1)} \ll 1$ already
breaks down at $k=k_{n=1}$. In order to avoid this problem, we
choose $k_* \gtrsim 10^2k_{s1}$, and try to find the likelihood
distribution of various values of $k_*$.  We also vary $k_{s1}$
and $p$ in addition to other cosmological parameters. The best fit
angular power spectrum in this case is given in curve~(b) of
Fig.~\ref{TT}.

The probability distribution of the likelihood values of $k_*$ and
$p$ are shown in Fig.~\ref{noncomtens}. This corresponds to the
mean likelihood analysis where the probability distribution of
$\chi^2$ is assumed to be Gaussian, as $P=e^{-(\chi-\chi_m)^2/2}$
(here $\chi_m$ is the minimum value of $\chi$). The best-fit
critical scale corresponds to
 \beq
 k_*=2.3 \times 10^{-2}\,{\rm
Mpc}^{-1}\,,
 \eeq
in which case the spectrum changes from (\ref{power2}) to
(\ref{powerspei}) around $100 \lesssim l \lesssim 200$.  The
distribution of $p$ is consistent with the result
of~\cite{Huang:2003zp}, obtained by using the WMAP data at the
scale $k=0.05\,{\rm Mpc}^{-1}$.  Other best-fit values are found
to be
 \beq
k_{s1}=3.24\times 10^{-6} {\rm Mpc}^{-1}\,,
 \eeq
$A_1=1.77\times 10^{-9}$, and
 \ba
&&\Omega_{\Lambda}=0.745\,,~ \Omega_b h^2=0.0209\,, ~\Omega_c
h^2=0.108\,,\nonumber \\&& \tau=0.0121\,,~ h=0.732\,.
 \ea
Since ${\cal P}_{\cal R}$ changes to scale-invariant for $k \le
k_*$, the angular power spectrum exhibits a better fit compared to
the standard case [compare curves~(b) and (c) in
Fig.~\ref{TT}]. This is an advantage of non-commutative inflation,
which allows for the running of the spectral index, due to the
existence of a cut-off momentum that arises from the stringy
uncertainty relation. Note that using only the formula
(\ref{power2}) leads to a power spectrum which is even more
suppressed for low multipoles, as seen in Fig.~\ref{TT} (a),
but this is ruled out since ${\cal P}_{\cal R}$ becomes negative.

We can constrain the string length scale, $L_s$, by using the
above best-fit values.  Since the two spectra (\ref{power2}) and
(\ref{powerspei}) are interpolated at $k_*$, we have
\beq \frac{L_s^{-1}}{M_{pl}} \simeq 2\pi \sqrt{\frac{2A_1}{p}}
\left(\frac{k_*}{k_0}\right)^{-1/(p-1)} \,, \label{Ls} \eeq
where we used the fact $k_{s1} \ll k_*$. The scale $k_0$ is chosen
as $k_0=0.05 \,{\rm Mpc}^{-1}$ in the numerical calculation of
CAMB. Substituting the most likely values of $k_*$, $p$ and $A_1$
in Eq.~(\ref{Ls}), one gets
\beq L_s^{-1} \simeq 10^{-4} M_{pl} \simeq 10^{14}\,{\rm GeV}\,,
\label{bestLs} \eeq
which corresponds to a string length scale, $L_s \sim
10^{-28}\,{\rm cm}$. This result must be interpreted with care. It
means that in the context of our postulated theoretical framework,
the best fit value of $L_s$ is the above one. From
Fig.~\ref{noncomtens} it is also clear that this value of $L_s$ is
more likely only by a modest amount than the value $L_s = 0$.

We also considered the case where the cosmologically relevant
scales are dominated by the spectrum (\ref{power2}), and obtained
a similar constraint to (\ref{bestLs}) by utilizing the likelihood
values of $k_{s1}$, $p$ and $A_1$.  Since $|-1/(p-1)| \ll 1$ in
Eq.~(\ref{Ls}), the change of $k_*$ around the scale $ k_* \sim
10^{-2}\,{\rm Mpc}^{-1}$ does not lead to any significant
modification for the estimation of $L_s$. The string length scale
is mainly determined by the ratio $A_1/p$. It is quite intriguing
that space-time non-commutativity opens up the possibility to
constrain the string scale by using the observational CMB data
sets.

The largest scales correspond to the initial time
with $\tau_k \sim kL_s^2 $.
Expanding the exact solution (\ref{tauexact})
around $\tau_k \sim kL_s^2 $, we get
\beq \label{powerspe3} {\cal P}_{\cal R} \simeq A_3
\left(\frac{k}{k_{s3}}\right)^{4/(p+1)}
\left[1+\left(\frac{k}{k_{s3}}\right)^{2/p}\right]^{-1}\,, \eeq
where
\ba
 \label{amp2}
A_3 &=& \frac{p}{4\pi^2}\!
\left[\!\frac{(k_0L_s)^{-4}}{5p(p+1)}\!\right]^{\!{p}/{(p+1)}}
\!\left(\!\frac{k_{s3}}{k_0}\!\right)^{\!{2}/{(p+1)}}
\!\left(\!\frac{k_0}{M_{pl}}\!\right)^{\!2\!}\\
k_{s3} &=& \left[\frac{4(p+1)^3}{5p}\right]^{p/2} (k_0L_s)^p
L_s^{-1}\,. \label{cut2} \ea
This is a blue tilted spectrum for $k \ll k_{s3}$. The critical
scale $k_{s3}$ is much less than $k_{s2}$; e.g., for $p=20$,
$k_{s3}\sim 10^{-5} k_{s2} \sim 10^{-11}k_{s1}$.
As long as the spectra are characterized by (\ref{power2}) and
(\ref{powerspei}) on the scales $10^{-4}\,{\rm Mpc}^{-1} \lesssim
k \lesssim 10^{-1}\,{\rm Mpc}^{-1}$, Eq.~(\ref{powerspe3}) is not
important, since it applies only well beyond $H_0^{-1}$, with
$k \ll 10^{-4}\,{\rm Mpc}^{-1}$. While the case (b) in
Fig.~\ref{TT} exhibits good agreement with the WMAP data, the
scale-invariant spectrum (\ref{powerspei}) is not sufficient to
explain significant loss of power around $l=2, 3$.  The amplitude of the
fluctuations tends to grow toward smaller multipoles due to the
Sachs-Wolfe effect even in the scale-invariant case.

Instead we can consider a case where the spectrum around $2 \le l
\lesssim 10$ is characterized by (\ref{powerspe3}). When $k \ll
k_{s3}$, Eq.~(\ref{powerspe3}) can be written in the form
\beq {\cal P}_{\cal R} \simeq A_3 \left\{ 1-\exp
\left[-\left({k \over
k_{s3}}\right)^{\alpha}\right] \right\}
\,,
\label{exp}
\eeq
with $\alpha=4/(p+1)$. This corresponds to the case of an
exponential cut-off in the spectrum analyzed
in~\cite{Contaldi,Cline,bfm}, except for the absence of the tilted
term, $k^{n_s-1}$.  In~\cite{Contaldi,Cline}, the value
$\alpha=3.35$ is chosen, and the TT spectrum is shown to be
suppressed for low multipoles. It is clear that $\alpha>1$ is
required for strong suppression on large scales (for example, see
Fig.~1 in~\cite{bfm} for $\alpha=1$ and $n_s=0.99$). Since $p$ is
larger than unity in our case, $\alpha$ for non-commutative
inflation is restricted to be smaller than 1. Therefore, we do not
have a significant suppression only around $l=2, 3$ as long as we use
the spectrum (\ref{powerspe3}) with $k \ll k_{s3}$.
Note, however, that the spectrum can be better fitted than the
standard $\Lambda$CDM model in power-law inflation.

\section{Discussion}

In this work we have found the best-fit parameters of a model of
inflation based on space-time
non-commutativity~\cite{Brandenberger:2002nq} when comparing to
the recent WMAP spectrum of CMB anisotropies.
The advantage of this approach is that one
uses stringy phenomenology to modify inflationary perturbations,
rather than imposing ad hoc modifications.
The stringy corrections ensure that the model is not subject to
the trans-Planckian problem of general inflationary models,
since the physical wavenumbers of modes have an upper bound.
At the same time, this feature means
that high-energy stringy effects modify the large-scale perturbations
rather than those on small scales,
since large-scale modes are generated outside the Hubble radius
and thus experience growth due to squeezing
for less time than they do for $L_s=0$. This model
automatically predicts that at large angular scales the spectrum
will be blue, thus providing a possible explanation for the
observed lack of power at the quadrupole and octopole. Roughly
speaking, requiring the correct location of the transition between
blue and red spectrum in our model determines the string length
scale, whereas the spectrum on smaller angular scales determines
the best fit value of the power-law exponent $p$.

Given that there are now a large number of possible explanations
of the observed deficit of power on large angular scales, it would
be of interest to look for special signals in our model not
present in the other proposed theoretical explanations. Work on
this subject is in progress. Note that our ``determination'' of
the length scale $L_s$ assumes that our class of models is in fact
correct, and that no secondary effects cause any deviations of the
spectrum.  The first point, in particular, is a major assumption
to be justified.

We have for simplicity assumed power-law inflation. In this case
$z=a\dot{\varphi}/H$ is proportional to $a$ due to the
time-independence of the small parameter, $\epsilon \equiv
\dot{\varphi}/H$. Therefore, the spectrum of curvature
perturbations is similar to that of gravitational waves, except
for their amplitudes.  The spectrum changes in general inflation
models because of the time variation of $\epsilon$.  In
particular, the running of the spectral index can be different
from the one in power-law inflation.  Although it may be in
general difficult to obtain the spectrum of primordial curvature
perturbations analytically, it will be interesting to extend our
analysis to other inflationary models by using a numerical
approach.  This would allow the exciting possibility to place more
generally applicable limits on the string length scale, in
addition to limits on the value of more general inflationary model
parameters.

Recently~\cite{Cremonini}, it was shown that a quantum deformation
of the wave equation on a cosmological background yields a
modified power spectrum analogous but not identical to
Eq.~(\ref{power2}). It would be of interest to constrain the
region of parameter space in those models through a likelihood
analysis similar to what we have done here, since this can provide
a powerful tool to pick up a possible trans-Planckian effect and
to distinguish between different string inspired cosmological
models.\\

{\bf Acknowledgements}

We are indebted to Antony Lewis for crucial advice and support in
implementing and interpreting the likelihood analysis. S.T. is
grateful to Bruce Bassett, Rob Crittenden and David Parkinson for
useful discussions. R.B. thanks George Efstathiou for important
advice. We also wish to acknowledge discussions with Stephon
Alexander at the beginning of this project. S.T. acknowledges
financial support from JSPS (No.~04942). R.M. is supported by
PPARC. R.B. is supported in part by the US Department of Energy
under Contract DE-FG02-91ER40688, TASK~A.




\begin{thebibliography}{99}

\bibitem{Guth}
A.~H.~Guth,
Phys.\ Rev.\ D {\bf 23}, 347 (1981); \\
K.~Sato,
Mon.\ Not.\ Roy.\ Astron.\ Soc.\  {\bf 195}, 467 (1981).

\bibitem{initial}
R.~H.~Brandenberger and J.~Martin,
Mod.\ Phys.\ Lett.\ A {\bf 16}, 999 (2001);\\
J.~Martin and R.~H.~Brandenberger,
Phys.\ Rev.\ D {\bf 63}, 123501 (2001).

\bibitem{review}
R.~H.~Brandenberger,
hep-th/0210186.

\bibitem{recent}
L.~Bergstrom and U.~H.~Danielsson,
JHEP {\bf 0212}, 038 (2002);\\
X.~Wang, B.~Feng and M.~Li,
arXiv:astro-ph/0209242;\\
V.~Bozza, M.~Giovannini and G.~Veneziano,
JCAP {\bf 0305}, 001 (2003);\\
C.~P.~Burgess, J.~M.~Cline, F.~Lemieux and R.~Holman,
JHEP {\bf 0302}, 048 (2003);\\
J.~Martin and R.~Brandenberger,
hep-th/0305161;\\
C.~P.~Burgess, J.~M.~Cline and R.~Holman,
hep-th/0306079;\\
O.~Elgaroy and S.~Hannestad,
astro-ph/0307011;\\
N.~Kaloper and M.~Kaplinghat,
hep-th/0307016.

\bibitem{Kogut:2003et}
A.~Kogut {\it et al.},
astro-ph/0302213.

\bibitem{Spergel:2003cb}
D.~N.~Spergel {\it et al.},
astro-ph/0302209.

\bibitem{Verde:2003ey}
L.~Verde {\it et al.},
astro-ph/0302218.

\bibitem{Peiris:2003ff}
H.~V.~Peiris {\it et al.},
astro-ph/0302225.

\bibitem{ge}
G.~Efstathiou, astro-ph/0306431.

\bibitem{low1}
S.~L.~Bridle, A.~M.~Lewis, J.~Weller and G.~Efstathiou,
astro-ph/0302306.

\bibitem{low2}
S.~de Deo, R.~R.~Caldwell and P.~J.~Steinhardt, Phys. Rev. D{\bf
67},
103509 (2003); \\
G. Efstathiou, astro-ph/0303127;\\
J.~P.~Uzan, A.~Riazuelo, R.~Lehoucq and J.~Weeks,
astro-ph/0303580.

\bibitem{Contaldi}
C.~R.~Contaldi, M.~Peloso, L.~Kofman and A.~Linde,
JCAP {\bf 0307}, 002 (2003).

\bibitem{Cline}
J.~M.~Cline, P.~Crotty and J.~Lesgourgues,
astro-ph/0304558.

\bibitem{Feng}
B.~Feng and X.~Zhang, astro-ph/0305020.

\bibitem{Yokoyama}
J.~Yokoyama,
Phys.\ Rev.\ D {\bf 59}, 107303 (1999).

\bibitem{KYY}
M.~Kawasaki, M.~Yamaguchi and J.~Yokoyama,
Phys.\ Rev.\ D {\bf 68}, 023508 (2003).


\bibitem{bfm}
M.~Bastero-Gil, K.~Freese and L.~Mersini-Houghton, hep-ph/0306289.

\bibitem{uncpr}
T.~Yoneya,
Mod.\ Phys.\ Lett.\ A {\bf 4}, 1587 (1989);\\
M.~Li and T.~Yoneya,
hep-th/9806240.


\bibitem{Alexander:2001dr}
S.~Alexander, R.~Brandenberger and J.~Magueijo,
Phys.\ Rev.\ D {\bf 67}, 081301 (2003).

\bibitem{Brandenberger:2002nq}
R.~Brandenberger and P.~M.~Ho,
Phys.\ Rev.\ D {\bf 66}, 023517 (2002).

\bibitem{excep}
Note that a different type of non-commutativity in inflation is
possible~\cite{fkm}, with qualitatively different features to
those considered here. Non-commutativity is applied on the sphere,
rather than the real spacetime non-commutativty considered here.
Modes are set to be absent at all times if they are inconsistent
with the non-commutativity constraint at the time of Hubble radius
crossing. By contrast, in our case modes are produced if they
initially do not satisfy the constraint.

\bibitem{fkm}
M.~Fukuma, Y.~Kono and A.~Miwa, hep-th/0307029.

\bibitem{com1}
The reason is that these modes undergo a shorter period of
squeezing than they do according to the standard calculations.

\bibitem{Huang:2003zp}
Q.~G.~Huang and M.~Li,
JHEP {\bf 0306}, 014 (2003).

\bibitem{MFB}
V.~F.~Mukhanov, H.~A.~Feldman and R.~H.~Brandenberger,
Phys.\ Rept.\  {\bf 215}, 203 (1992).

\bibitem{CAMB}
http://camb.info/

\bibitem{antony} A.~Lewis, A.~Challinor and A.~Lasenby,
Astrophys.\ J.\  {\bf 538}, 473 (2000);\\
A.~Lewis and S.~Bridle,
Phys.\ Rev.\ D {\bf 66}, 103511 (2002).

\bibitem{WMAP}
http://lambda.gsfc.nasa.gov/

\bibitem{CBI}
T.~Pearson {\it et al.}, astro-ph/0205388.

\bibitem{VSA}
K.~Grainge {\it et al.}, astro-ph/0212495.

\bibitem{ACBAR}
C.~L.~Kuo {\it et al.}, astro-ph/0212289.


\bibitem{Cremonini}
S.~Cremonini,
hep-th/0305244.

\end{thebibliography}
\end{document}